\documentclass[jcp,twocolumn]{revtex4-1}
\usepackage{amsmath}
\usepackage[english]{babel}
\usepackage{latexsym}
\usepackage{mathrsfs}
\usepackage{graphicx}
\usepackage{graphics}
\usepackage{color}
\usepackage{bm}
\usepackage{units,nicefrac}

\begin{document}

\title{Extended Short-Range Order Determines the Overall Structure of Liquid Gallium}

\author{Anatolii~V.~Mokshin} \email{anatolii.mokshin@mail.ru}
\affiliation{Kazan Federal University, Kazan, 420008 Russia}

\author{Ramil~M.~Khusnutdinoff} \email{khrm@mail.ru}
\affiliation{Kazan Federal University, Kazan, 420008 Russia}

\author{Bulat N. Galimzyanov} \email{bulatgnmail@gmail.com}
\affiliation{Kazan Federal University, Kazan, 420008 Russia}

\author{Vadim~V.~Brazhkin}
\affiliation{Institute for High Pressure Physics, RAS, 108840 Moscow, Russia}

\begin{abstract} 
	Polyvalent metal melts (gallium, tin, bismuth, etc.) have microscopic structural features, which are  detected by neutron and X-ray diffraction and which are absent in simple liquids. Based on neutron and X-ray diffraction data and results of \textit{ab initio} molecular dynamics simulations for liquid gallium, we examine the structure of this liquid metal at atomistic level. Time-resolved cluster analysis allows one to reveal that the short-range structural order in liquid gallium is determined by a range of the correlation lengths. This analysis performed over set of independent samples corresponding to equilibrium liquid phase  discloses that there are no stable crystalline domains as well as molecule-like Ga$_2$ dimers typical for crystal phases of gallium. Structure of liquid gallium can be reproduced by the simplified model of the close-packed system of soft quasi-spheres. The results can be applied to analyze the fine structure of other polyvalent liquid metals.
\end{abstract}

\maketitle

\section{Introduction}

Pure metal melts consisting only of atoms of same type usually
refer to simple liquids \cite{Dyre_PRX_2016}. In these metal melts, the
ion-ion interactions screened by an electron gas can be reproduced, as
expected, in a reasonable approximation by a spherically symmetric potential
$u(r)$. These expectations are partially justified. Indeed, in liquid alkali metals,
the pairwise additive pseudo-potentials yield more correct results for the
and collective particle dynamics as compared to the results
obtained with the more sophisticated
EAM-potentials~\cite{Gonzalez_2001,Mokshin/Galimzyanov_JCP_2018,Khusnutdinov/Mokshin/Galimzyanov_JETP_2018}.
With an increase in valence, the situation becomes more complicated. As
detected by means of neutron and X-ray diffraction~\cite{Mokshin/Khusnutdinoff_JETP_2015,Li_2014,IXS_ab_init,Demmel_2020}, the structure of liquid gallium, which belongs to the boron group in the periodic table of the chemical elements, differs from the structure typical for simple liquids.
First of all, the reduced positions of the peaks of the radial distribution
function $g(r)$ and of the static structure factor $S(k)$ deviate
significantly from the values which should be in the case of a simple liquid. Here, $r$ and $k$ are the distance and the wave number, respectively. The main maximum of the static structure factor $S(k)$ of liquid gallium is highly
asymmetric, and with approaching the melting point a pronounced shoulder is formed in its left side. These structural features do not appear for the case of simple liquids, but are also detected for other pure metal melts, namely, Ge, Sn, As, Sb, and Bi. In this study, we shall show that the structural features of liquid gallium can be explained by a simple yet detailed analysis of neutron/X-ray diffraction data in combination with \textit{ab initio} molecular dynamics (AIMD) simulation results. [Simulation details are given in Supplementary Information]. For this, the thermodynamic states corresponding to equilibrium liquid along the isobar $p=1.0$~atm will be considered.

\begin{figure}[h]
	\centering
	\includegraphics[width=8cm]{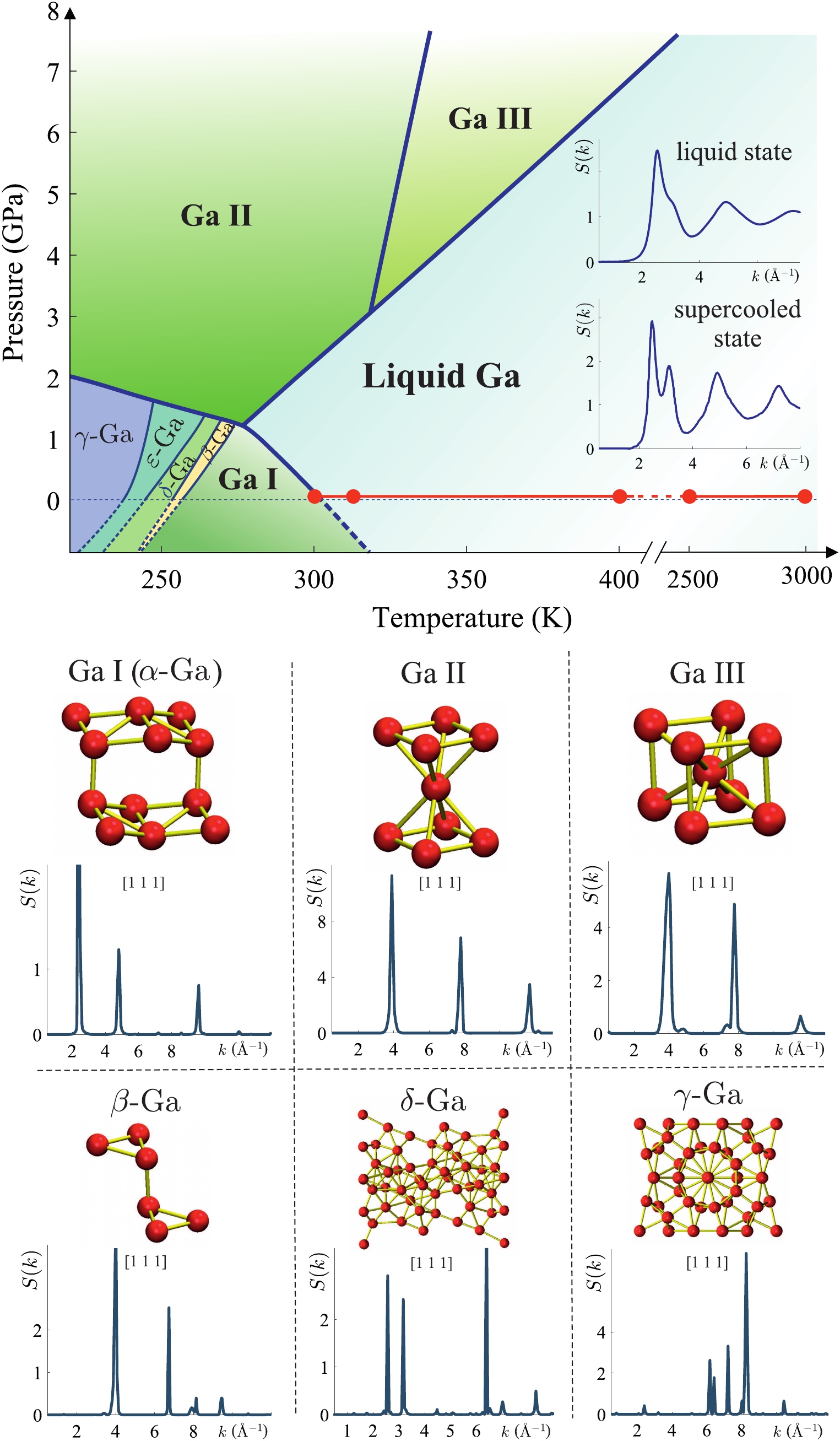}
	\caption{(Color online) \textbf{Top panel:} Phase diagram of gallium on the basis of the data from Refs.~\cite{Bosio_cryst_Ga_II_III,Cicco_PRB_2001}; red filled circles indicate the liquid phase states considered in this study.    \textbf{Bottom panel:} Crystal structures and corresponding static structure factors estimated for the direction [$1\,1\,1$].}
	\label{fig01}
\end{figure}

\section{Microscopic structure of liquid gallium}

\textbf{Relevant information about gallium.} -- Gallium is characterized by physical properties typical for a metal and it has remarkable phase diagram~\cite{Brazhk,Shulte,Bosio_cryst_Ga_II_III} (see Fig.~\ref{fig01}). Equilibrium liquid phase occupies a significant part of the experimentally available range of the phase diagram~\cite{Li_2011}. For the isobar $p=1.0$~atm, the melting temperature is $T_m= 302.93$~K, whereas the boiling temperature is $T_b=2\,477$~K \cite{Iida_1993,Brazhkin_JETP2008}, and, therefore, the range of equilibrium liquid phase is more than $2\,000$~degrees. It is well known by now that the phase diagram of gallium comprises at least seven crystal phases~\cite{Li_2011}. The crystal phase with orthorhombic structure, known as Ga\,I or $\alpha$-Ga, is thermodynamically stable at relatively low and moderate pressures. In particular, on the isobar $p=1.0$~atm, only this crystal phase is detectable as stable. Eight atoms form a unit cell, whose lattice parameters are $a=4.5192$~\AA, $b=7.6586$~\AA, $c=4.5258$~\AA; and $\alpha$=$\beta$=$\gamma$=$90^\circ$. Six neighbors of an arbitrary atom are arranged in pairs at the characteristic distances $2.70$~\AA, $2.73$~\AA  \, and $2.79$~\AA; and the atom and its six neighbors are located within a layer of the width $1.49$~\AA. The last (seventh) neighbor of selected atom is the closest one and is at the distance $r_b=2.46$~\AA. As originally assumed and then confirmed experimentally, such the pairs of atoms form molecule-like Ga$_2$ dimers~\cite{Zuger,Walko}.

Two other stable crystal phases -- tetragonal phase Ga\,II and orthorhombic phase Ga\,III -- appear only at high pressures, whereas the remaining four ($\beta$-, $\gamma$-, $\delta$-, $\epsilon$-) modified solid state phases of gallium are not stable. Another remarkable property of gallium is ability under sufficiently rapid cooling of the melt to generate an amorphous solid phase with a high degree of structural disorder~\cite{Chen_1997}. This property is rarely found in one-component (pure) metals and may be due to the specific structure of liquid gallium. The splitting of the main maximum of the static structure factor that is observed for the supercooled liquid gallium can be considered as manifestation of the characteristic correlation lengths associated with the short-range order~(see inset in Fig.~\ref{fig01}).

\begin{figure*}
	\centering
	\includegraphics[width=18cm]{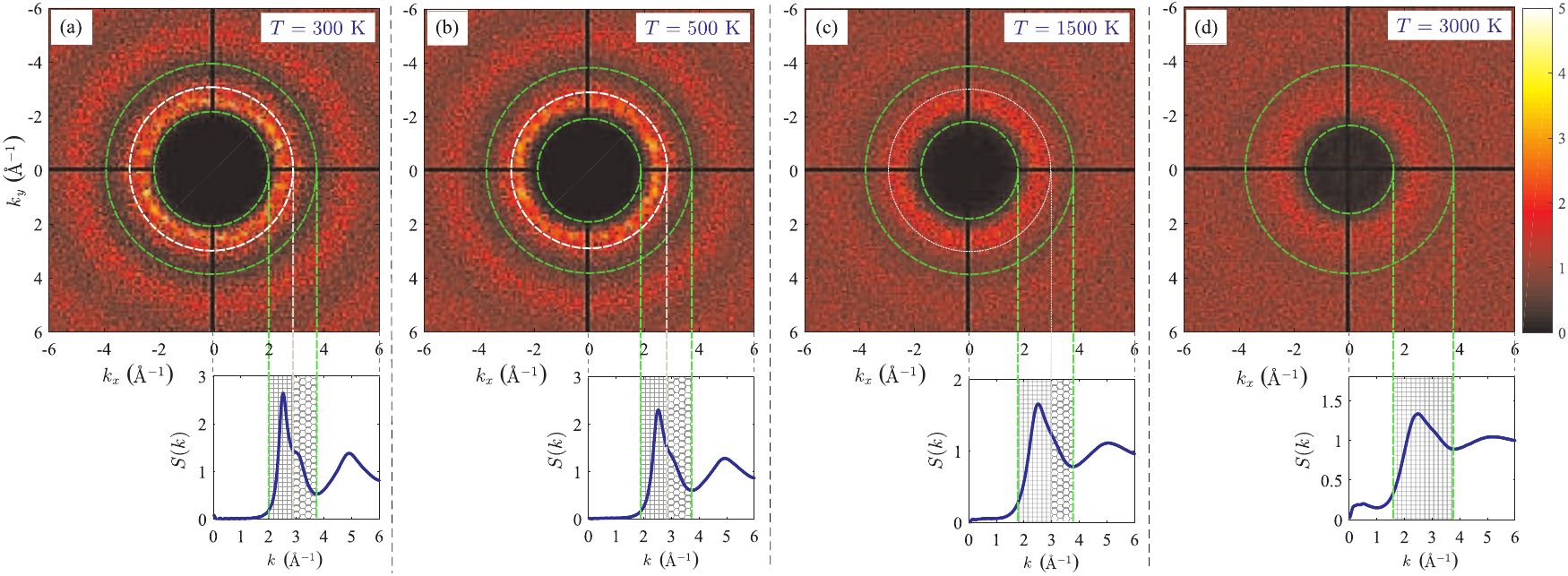}
	\caption{(Color online) Diffraction pattern $S(k_x,k_y)$ and corresponding static structure factor $S(k)$ of liquid gallium evaluated for the equilibrium thermodynamic states with the pressure $p=1.0$~atm and the temperatures $T=300$, $500$, $1\,500$ and $3\,000$~K by means of AIMD simulations. Panels (a)-(c) correspond to the liquid states, and panel (d) presents results for the dense vapour state. Ring enclosed between wave number values $k=1.9$~\AA$^{-1}$ and $3.8$~\AA$^{-1}$, where $k=\sqrt{k_x^2+k_y^2}$, corresponds to $k$-values of the main maximum of the static structure factor $S(k)$. Heterogeneity of $S(k_x,k_y)$ inside this ring in the radial direction is clearly manifested at temperatures $T=300$~K and $500$~K and is evidence of the asymmetry of the main peak of the static structure factor.}
	\label{fig02}
\end{figure*}

\textbf{Structure of liquid gallium: Brief overview of current understanding.} -- As mentioned above, a shoulder on the right (high-$k$) side of the main peak of the static structure factor $S(k)$ was initially revealed by means of neutron and X-ray scattering~\cite{Li_2014,Narten_1972,Blagov_2014,Li_Sci_Rep_2017}. Subsequent molecular dynamics simulations confirmed the presence of this feature of the static structure factor~\cite{IXS_ab_init,Yang_2011}. Namely, when we move along the isobar $p=1.0$~atm from high temperatures to the melting temperature $T_m$, then the first signs of shoulder in $S(k)$ appear at the temperature about $800$~K. With a further decrease in temperature, this shoulder becomes more pronounced. This is seen in Fig.~\ref{fig02}, where we show the diffraction pattern $S(k_x,k_y)$ and the corresponding static structure factor $S(k)$ of liquid gallium evaluated for different temperatures by means of AIMD simulations~\cite{Acta_M_2019}.

The diffraction experiments provide only an averaged information about the structure of a studied system, and some microscopic structural details may remain undetected. Unlike this, \textit{ab initio} molecular dynamics simulations would shed light on the problem. Surprisingly, despite the fact that AIMD simulations of liquid gallium were performed by different research groups~\cite{Cicco_EPL_1994,IXS_ab_init,Gaston_2014,Klein_2018}, there is still no convincing and generally accepted explanation for the physical origin of the features observed in the contour line of $S(k)$. Currently, there are known three main points of view proposed to explain these  features. Namely, it is assumed that this asymmetry as well as the shoulder in $S(k)$ might be caused by (i) ensemble of locally ordered structures (clusters, `four-atom units')~\cite{Tsay_PRB1994,Chen_2016}, or (ii) ensemble of covalent-bonded Ga$_2$ dimers~\cite{Gong_1993}, or (iii) specific interparticle interaction, which includes the ledge-shaped repulsive core and the long-ranged Friedel oscillations~\cite{Hafner,Tsai_2010}. Having available X-ray and neutron diffraction data supplemented by our results of AIMD simulations, we are aimed in this study to analyze in detail the microscopic structure of liquid gallium as well as to revise critically the existing hypotheses explaining the features of its structure.

\textbf{Direct mathematical interpretation of the diffraction pattern.} -- In the case of disordered systems such as liquids and amorphous solids, there are no dedicated directions in the structure and, therefore, the static structure factor $S(k)$ is related to the radial distribution function $g(r)$ as follows
\begin{eqnarray}
\label{eq: S_k_radial} S(k) &=& 1 + \frac{4 \pi
	\rho}{k}\int_{0}^{\infty} r [g(r) - 1] \sin (kr) dr  \\
&=& 1 - \frac{4 \pi \rho}{k} \frac{d}{dk}\int_{0}^{\infty} [g(r) -
1] \cos (kr) dr,\nonumber
\end{eqnarray}
where $\rho$ is the number density. Let $\hat{g}(k)$ be the `image' of the `original' function $g(r)$ resulted from the integral transform
\begin{equation} \label{eq: image}
\hat{g}(k) = \int_0^{\infty} g(r) p(kr) dr,
\end{equation}
where $p(\ldots)$ is a periodic function (sine function, cosine function). Then, from a mathematical point of view, the peaks positions of the function $S(k)$, which is conjugated directly with $\hat{g}(k)$ as seen from Eqs.~(\ref{eq: S_k_radial}) and (\ref{eq: image}), should provide the information about the characteristic correlation lengths and/or the inherent periods in the `original' function $g(r)$. As an example, let $k_m$ be the location of the main peak of the structure factor $S(k)$ in the case of a dense single-component atomistic system. Then, the characteristic correlation length $r_m=(2\pi)/k_m$ will give an approximate estimate of the average linear size $r_0$ of an atom in this system. In turn, the quantity $r_0$ can be simply defined from the radial distribution function $g(r)$: the quantity $r_0$ sets the location of the first (main) peak of the radial distribution function $g(r)$ on the $r$-axis. Nevertheless, the more pronounced oscillations of the function $g(r)$, the more the correlation length $r_m$ underestimates correct value of $r_0$.
For example, for simple dense liquids near the melting temperatures, the function $g(r)$ is explicitly oscillating, and, instead of a direct equivalence between $r_m$ and $r_0$, the following universal relation holds (see Ref.~\cite{JETP_Lett_2017_en}):
\begin{equation} \label{eq: ratio}
r_m \equiv (2\pi)/k_m \simeq 0.85 \; r_0.
\end{equation}
We note, finally, that the peaks of $S(k)$ at higher wave numbers, i.e. at $k>k_m$, arise due to oscillations in the radial distribution function $g(r)$ and, therefore, due to quasi-periodicity of arrangement of the pseudo-coordination spheres.
\begin{figure}
	\centering
	\includegraphics[width=7cm]{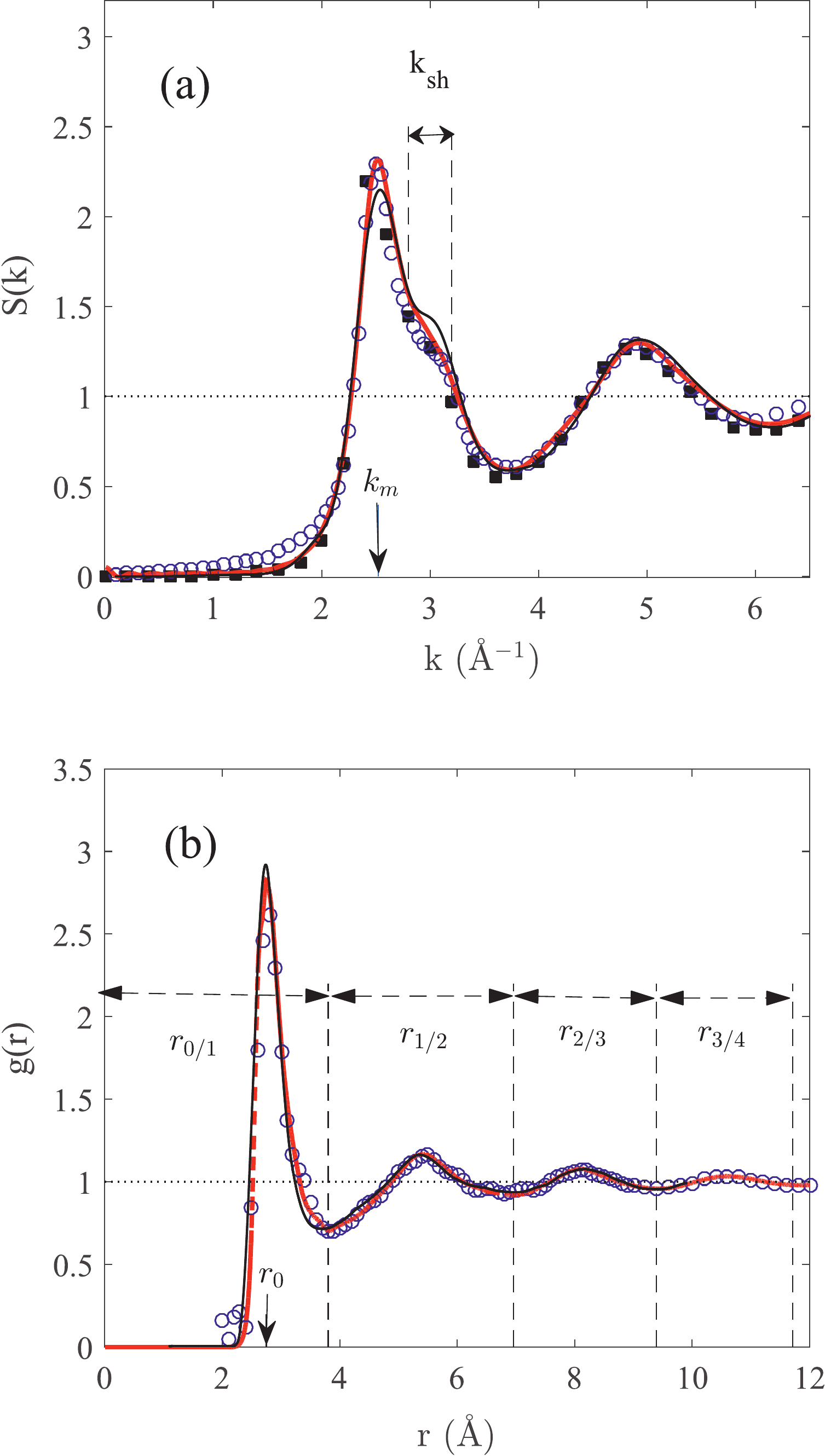}
	\caption{Static structure factor $S(k)$ and radial distribution function $g(r)$ of liquid gallium near melting: circles present X-ray diffraction data for the state with the temperature $T=323$~K~\cite{Waseda}, squares are neutron diffraction data for melt with $T=326$~K~\cite{Bellissent_1989}, dashed (red) line represents MD simulation results at the temperature $T=313$~K~\cite{Mokshin/Khusnutdinoff_JETP_2015} and solid (black) line shows \textit{ab initio} MD simulation results at the temperature $T=325$~K. In \textbf{panel (a)}, the wave number $k_m=2.52$\,\AA$^{-1}$ is the location of the main maximum of $S(k)$, and $k_{sh} = [2.8,\,3.2]$\,\AA$^{-1}$ is the wave number range of the high-$k$ shoulder. In \textbf{panel (b)}, the length $r_0=2.75$\,\AA \ corresponds to the location of the first peak in $g(r)$; the second, third and fourth peaks of $g(r)$ are located at $r=5.404$, $8.114$ and $10.59$\,\AA, respectively; whereas the size of the first, second and third pseudo-coordination shells are $r_{0/1}=3.803$\,\AA, $r_{1/2}=3.159$\,\AA, $r_{2/3}=2.43$\,\AA \ and $r_{3/4}=2.318$\,\AA, respectively, as evaluated from the location of the corresponding minima.}
	\label{fig03}
\end{figure}

The above considerations can now  be used to analyse the diffraction data for liquid gallium near melting. In Fig.~\ref{fig03}, we present the static structure factor $S(k)$ and the radial distribution function $g(r)$ as obtained from X-ray and neutron diffraction~\cite{Waseda,Bellissent_1989}, classical molecular dynamics simulations with the EAM-potential~\cite{Mokshin/Khusnutdinoff_JETP_2015} and \textit{ab initio} molecular dynamics simulations.
First of all, one finds the following correspondence between the correlation length $r_m=(2\pi)/k_m$ and the average linear size $r_0$ of gallium atom:  $r_m \simeq 0.91 \; r_0$ (see Fig.~\ref{fig03} and the figure caption), which sets the larger ratio between both the lengths, than this is for the  simple liquids [see Eq.(\ref{eq: ratio})]. Mathematically, this means that the maxima of the distribution function $g(r)$ should be located closer to each other in the case of liquid gallium compared with the case of simple liquids. This \textit{can be} interpreted as evidence of the medium-range structural order in liquid gallium that is in agreement with some conclusions of Ref.~\cite{Tsay_PRB1994,Chen_2016}. Nevertheless, it is still unclear whether this order is due to the presence of locally ordered structures. Further, as seen in Fig.~\ref{fig03}(a), the shoulder of $S(k)$, which is located within the wave number range $k \in [2.8,\,3.2]$\,\AA$^{-1}$, is manifestation of the spatial correlations with the characteristic lengths  $r \in [1.96,\, 2.24]$\,\AA. Taking into account that the first (main) peak of the radial distribution function is located within the range  $r\in[2.2,\,3.8]$\,\AA, one concludes that the shoulder in $S(k)$ \textit{can} occur due to extremely short particle bonds. Thus, both the factors -- the medium-range structural order and the covalent-like Ga$_2$ dimers -- can be the causes of the structural anomalies of liquid gallium found in the diffraction patterns. On the other hand, the crystal-like clusters and the covalently bonded dimers must be characterized by long lifetimes (at least, compared to the time of structural relaxation), and the presence of these clusters and/or dimers can be verified using the time-resolved cluster analysis of the structural data.
\begin{figure}
	\centering
	\includegraphics[width=9cm]{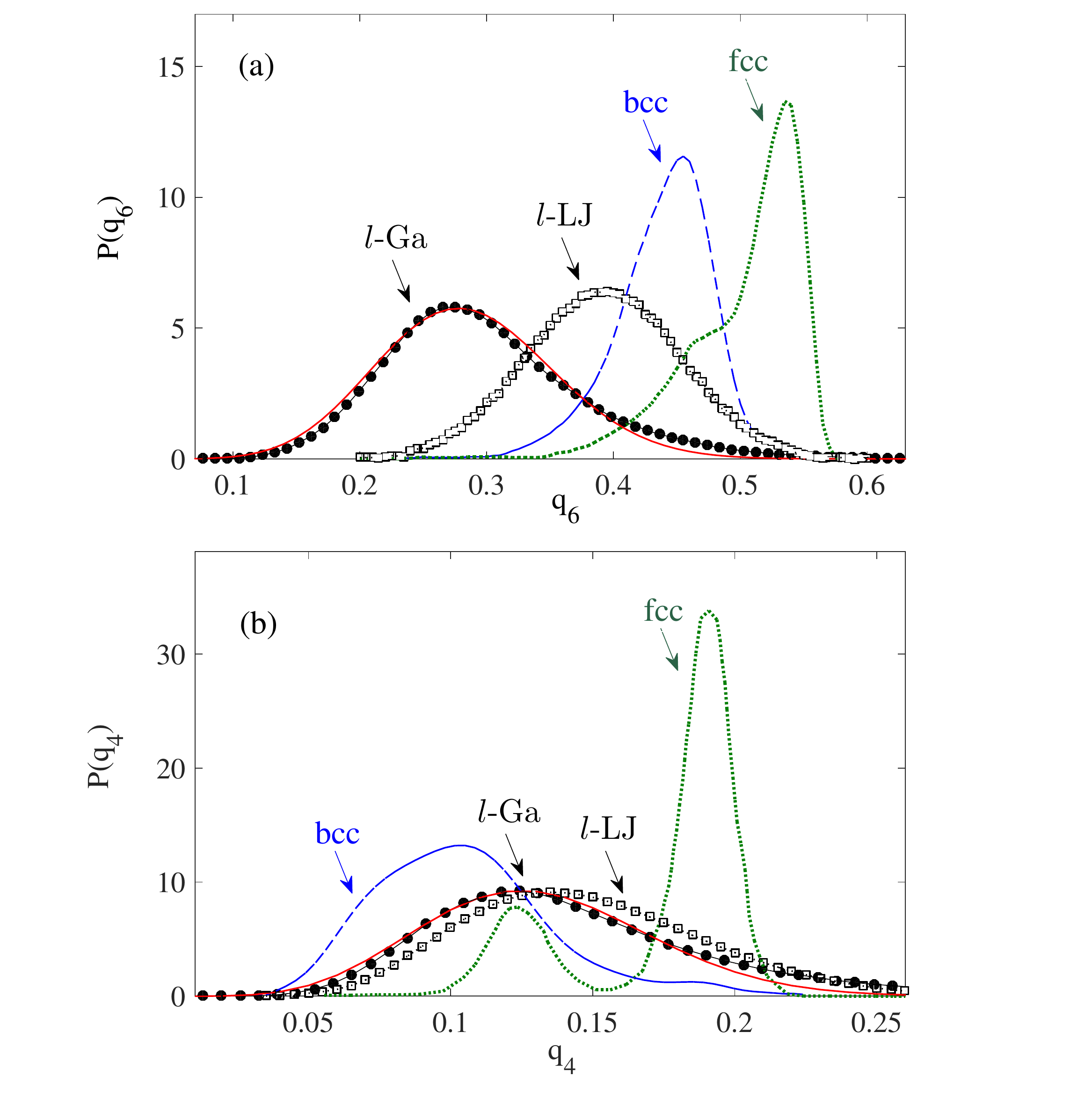}
	\caption{Distributions of the $q_6$- and $q_4$- local order parameters  evaluated for liquid gallium  at the temperature $T=325$\,K  (filled circles), for Lennard-Jones liquid (squares) as representative simple liquid, for crystalline solids with bcc-lattice (dashed line) and fcc-lattice (dotted line). Results for Lennard-Jones liquid is taken from Ref.~\cite{Mokshin/Khusnutdinoff_JETP_2015}.}
	\label{fig04}
\end{figure}

\textbf{Cluster analysis results, local orientational order and neighborhood time.} -- The local structure of the system can be examined by means of the local orientational order parameters
\begin{equation}
q_{l}(i) = \left ( \frac{4\pi}{2l+1} \sum_{m=-l}^{l} \left | \frac{1}{N_b(i)} \sum_{j=1}^{N_b(i)} Y_{lm} (\hat{\mathbf{r}}_{ij})  \right |^2 \right )^{1/2}
\label{eq: order_parameter}
\end{equation}
evaluated for an each $i$th particle. Here, the summation is only over all $N_b(i)$ neighbor particles located inside the first pseudo-coordination shell; $\hat{\mathbf{r}}_{ij} = \mathbf{r}_{ij}/|\mathbf{r}_{ij}|$ is the unit vector that sets the direction from particle $i$ to its neighbor $j$; and even-$l$ spherical harmonics have to be considered only~\cite{Steinhardt}. In the case of a three-dimensional monatomic high density system,  the local structural order can be of four- and/or six-fold orientational symmetry, and, therefore, we need to evaluate only the local order parameters $q_4$ and $q_6$. The contour lines of the distributions of the local orientational order parameters differ for crystalline phases as well as for disordered (liquid) phase (see Fig.~1 in Ref.~\cite{Wolde} and related discussion).

Both the distributions of these local order parameters, $P(q_6)$ and $P(q_4)$, evaluated from AIMD simulation data for liquid gallium near melting are unimodal and have contour lines typical for simple liquids (see Fig.~\ref{fig04}). Namely, these distributions are reproduced accurately by the model law:
\begin{equation}
P(q_l) = (A_l q_l)^{n_l} \mathrm{e}^{-C_l(q_l-q_l^{\mathrm{max}})^2},\ \ \ \ \ l=4,\;6,
\label{eq: ql}
\end{equation}
with the following parameters of the distributions $P(q_6)$ and $P(q_4)$: $A_4=20.6$, $n_4=4.4$, $C_4=160$, $q_4^{\mathrm{max}}=0.015$, $A_6=6.5$, $n_6=5.7$, $C_6=65$, $q_6^{\mathrm{max}}=0.12$. The results of cluster analysis obtained for liquid gallium are completely different from the case when a system contains crystalline domains with fcc- and bcc-lattice. Further, as follows from the low-$q$ tail of the distribution $P(q_4)$ and from the high-$q$ tail of the distribution $P(q_6)$ obtained for liquid gallium and presented in Fig.~\ref{fig04}, there are the particles in the system, which can be formally identified as incoming in `local ordered structures'. Nevertheless, the fraction of these particles is even lesser than as in the case of simple equilibrium Lennard-Jones liquid as it follows from $P(q_4)$ and $P(q_6)$. Thus, the equivalence of the given results to the case of equilibrium Lennard-Jones liquid can be considered as evidence of the absence of crystalline-like clusters in liquid gallium.

\begin{figure}
	\centering
	\includegraphics[width=10cm]{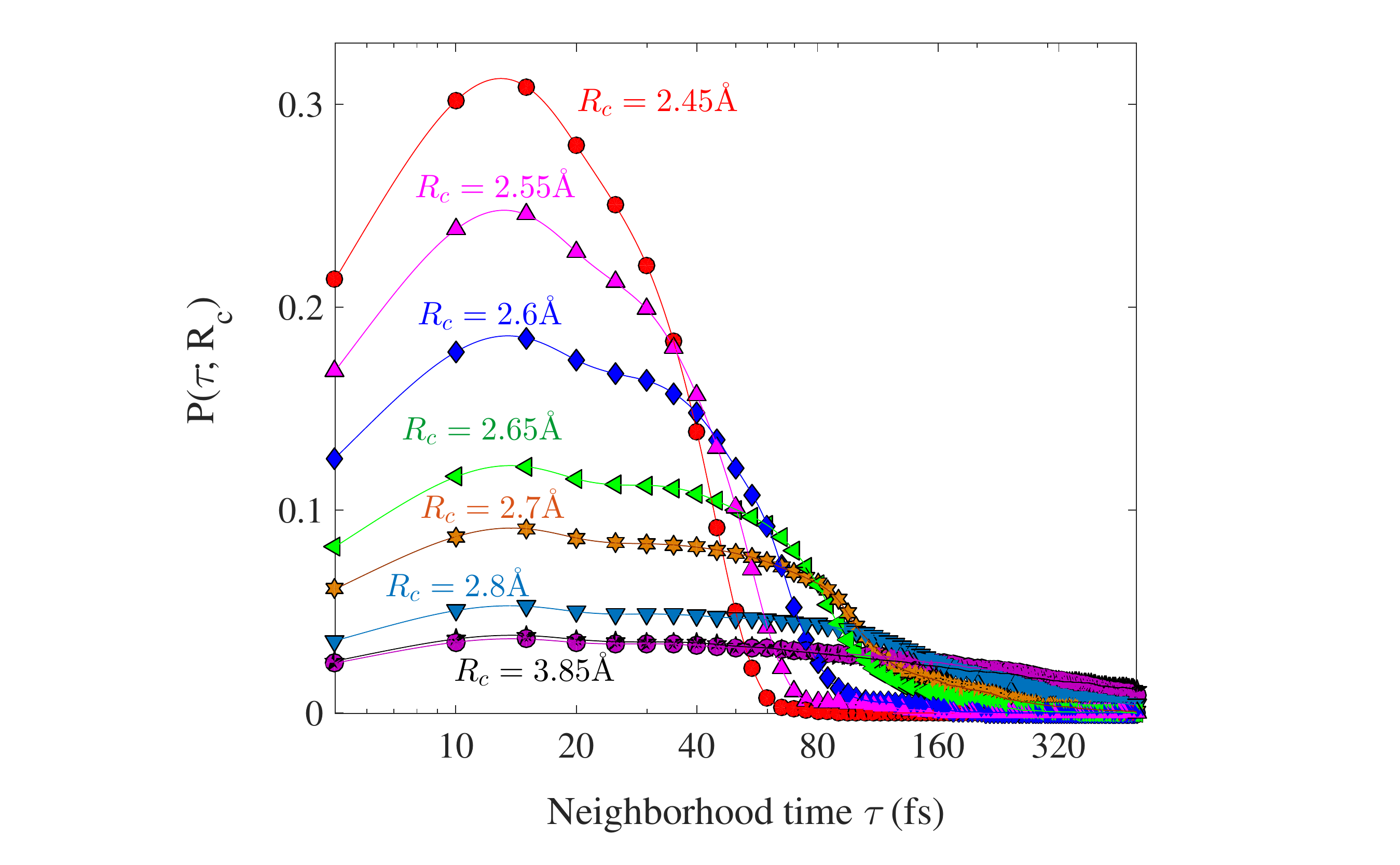}
	\caption{Distributions of the neighborhood time $\tau$ of two atoms when the distance between them is less than $R_c$. The cutoff distance $R_c$ sets the spatial scale within the first pseudo-coordination shell and it was taken as  $R_c = [2.45,\, 2.55,\, 2.6,\, 2.65,\, 2.7,\, 2.8,\, 3.0,\, 3.85]$\,\AA. The temperature of the equilibrium melt is $T=325$\,K. The most probable value of this time is determined by location of the distribution maximum and it is assessed at $13.0 \pm 0.4$~fs. Note that the x-axis of this plot is given on a logarithmic scale.
	}\label{fig05}
\end{figure}

Now, having cluster analysis results of AIMD simulations, we can revise the hypothesis of dimers Ga$_2$ in gallium melt.  If a pair of atoms forms a covalent bond, then the lifetime of such a covalent bond should be much larger than the time scale of the thermal motion of an atom. Let us define the time of the neighborhood $\tau(R_c)$ of two particles as the time when these particles are at a distance not exceeding a threshold value $R_c$. Taking into account that the covalent bond of Ga$_2$ dimers in crystalline phase is $r_b=2.46$~ \AA, it is advisable to choose a value of $R_c$ comparable with $r_b$.

In Fig.~\ref{fig05}, the distributions of the neighborhood time, $P(\tau)$, computed at various values of the cutoff distance $R_c$ for the thermodynamic state near melting are presented. All the distributions have a pronounced maximum setting the most probable neighborhood time $\tau=13.0 \pm 0.4$~fs. Given that the thermal velocity of gallium atom at this temperature is $v_T = 341$\,m/s, it can be found that the most probable neighborhood time $\sim 13.0$~fs corresponds to the collision time of a pair of atoms.  As follows from the distributions at the cutoff distances up to $R_c=2.6$\,\AA, that is larger than the covalent bond $r_b=2.46$\,\AA \ for   $\alpha$-Ga crystalline phase, these distributions, $P(\tau,R_c\leq2.6\,$\AA),  decay completely over the time scale $0.1$\,ps. In particular, the largest neighborhood time of two particles within the range of the linear size  $R_c=2.6$\,\AA \  is  $0.1$\,ps. These timescales are too short to be associated with the lifetimes of covalent bonds. The distribution $P(\tau,R_c=3.85\,$\AA), which corresponds to the particle dynamics on the spatial scale slightly larger than the size of the first pseudo-coordination shell [refer to Fig.~\ref{fig03}(b)], decays over the time scale $\sim 0.7$\,ps. Note that the time scale $0.7$\,ps is the largest one revealed for the dynamics of neighboring particles, and this time is \textit{typical} for \textit{thermal diffusion motion} of the particles. Consequently, there are no neighboring particles in the system that can be identified as bound. This indicates absence in the system of covalently bound dimers and any stable local structures, clusters.

\textbf{The shortest correlation lengths: Liquid gallium \textit{vs.} polydisperse system of hard spheres.} -- Another simple, but convenient way to evaluate the most short correlation lengths in the structure of liquid gallium is to match its structure with the structure of a close-packed polydisperse system of hard spheres, which has the same radial distribution function. Namely, peaks in the particle size distribution $P(\sigma)$ of the corresponding polydisperse system will manifest the shortest correlation lengths of the original system.

\begin{figure}
	\centering
	\includegraphics[width=9cm]{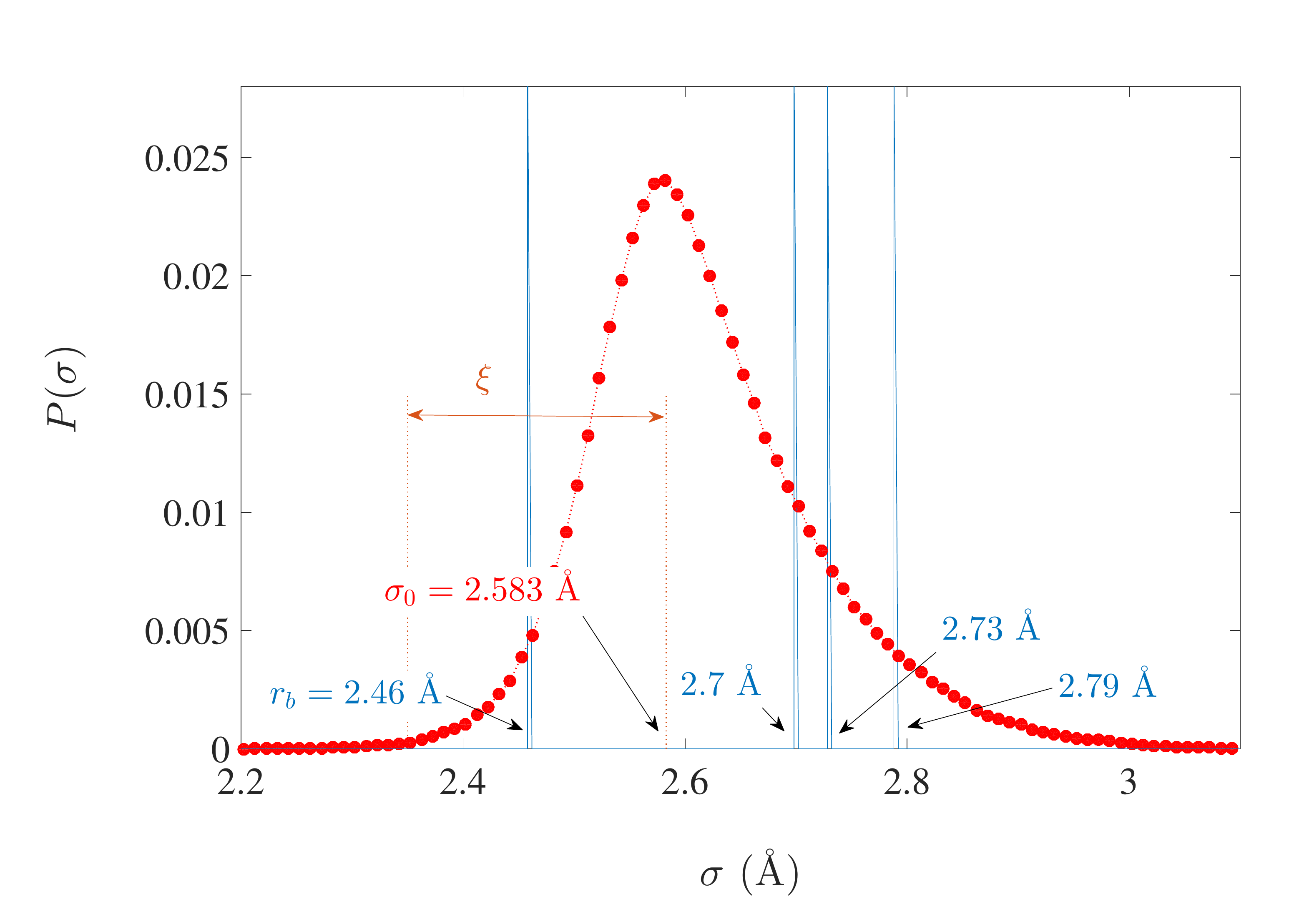}
	\caption{ (Color online)
		Particle size distribution function $P(\sigma)$ of a close-packed polydisperse system of hard spheres, the structure of which is given by the same structure functions $S(k)$ and $g(r)$ as for liquid gallium at $T=325$\,K in Fig.~\ref{fig03}(b). The maximum of $P(\sigma)$ is located at $\sigma_0=2.583$\,\AA.
		Vertical (blue) lines present size distribution function of a close-packed polydisperse system, which reproduces the structure of crystalline $\alpha$-Ga phase at the temperature $T=0$\,K.
	}\label{fig06}
\end{figure}

Taking AIMD simulation data with the defined positions of gallium atoms for an instantaneous configuration, we matched the close-packed polydisperse system of hard spheres, whose the radial distribution function $g(r)$ is identical to this function of liquid gallium. The polydisperse system was constructed as follows. Initially, the particles of an infinitesimal size were placed in the positions of gallium atoms. Then, a random particle was selected, and its size was increased by the small increment $\Delta r=0.003$\,\AA. If the particle touched a neighboring particle, then its size remained unchanged. The procedure was performed cyclically, until a close-packed system was formed. We note that this procedure was performed for more than thousand liquid gallium configuration data, and the particle size distribution $P(\sigma)$ of the model polydisperse system of hard spheres was determined as an average result.

It turns out that the structure of liquid gallium at the temperature $T=325$\,K is equivalent to the polydisperse system, the particle size distribution $P(\sigma)$ of which is given by the function shown in Fig.~\ref{fig06}.
The distribution $P(\sigma)$ is continuous and located on the length range $\sigma \in [2.35,\;3.01]$\,\AA, that exactly corresponds to the distances of the left half of the first maximum of the radial distribution function $g(r)$ for liquid gallium [see Fig.~\ref{fig03}(b)]. The single well-defined peak of the distribution is at $\sigma_0=2.58$\,\AA, that defines the lower limit value of the correlation length associated with the short-range order in gallium. Note that the length $\sigma_0$ is larger than the covalent bond $r_b$, and there are no peculiarities in the distribution $P(\sigma)$ at values of $\sigma$ comparable with $r_b$. This indicates the absence of dimers with a short bond.

\begin{figure*}
	\centering
	\includegraphics[width=14cm]{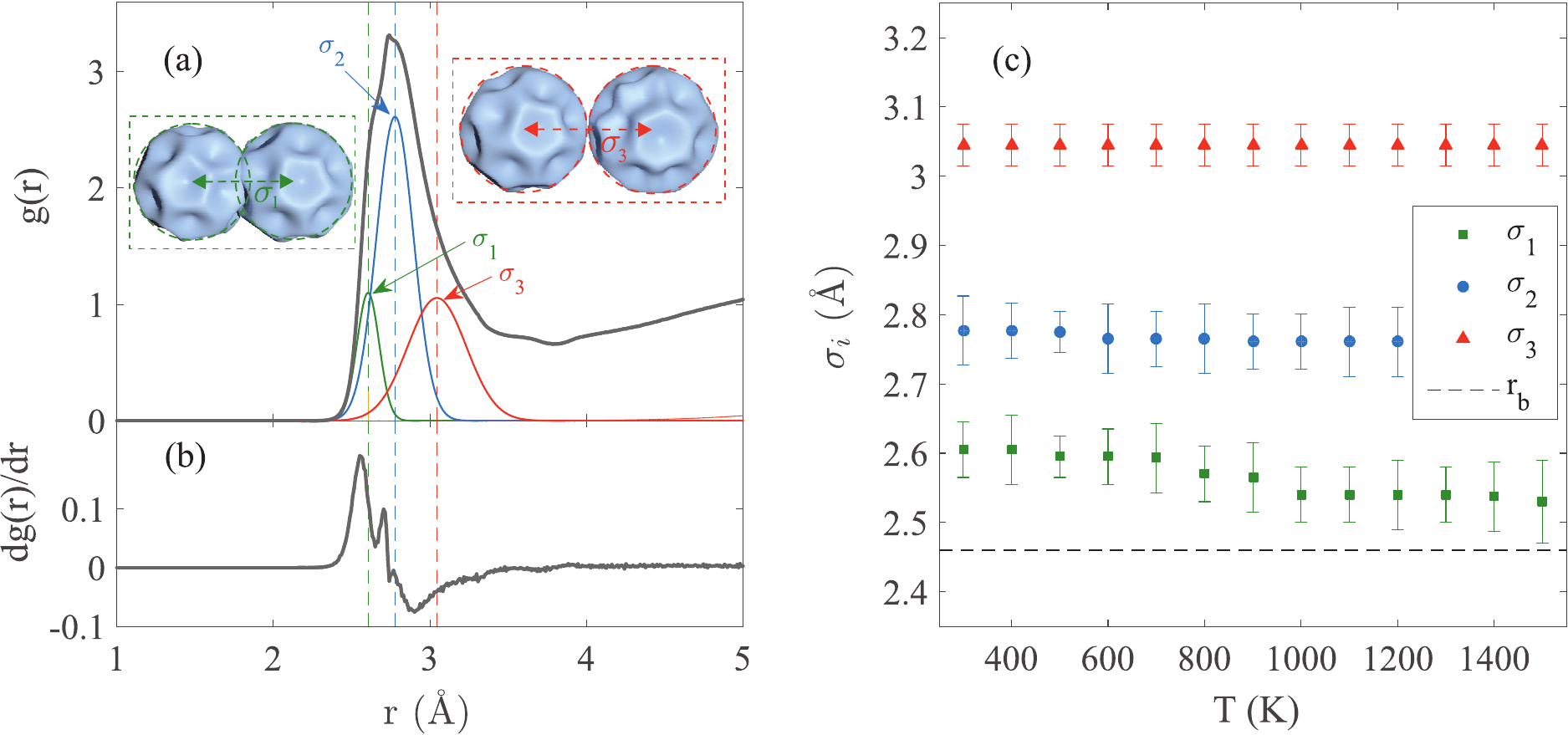}
	\caption{(Color online) (a) Radial distribution function $g(r)$ and (b) the first derivative of this function, $dg(r)/dr$, evaluated for liquid gallium at the temperature $T=325$~K. Vertical dashed lines indicate the correlation lengths $\sigma_1$, $\sigma_2$ and $\sigma_3$, which define the positions of the Gaussians used to fit the first peak of $g(r)$. (c) Correlation lengths $\sigma_1$, $\sigma_2$ and $\sigma_3$ vs. temperature. Horizontal dashed line indicates the length of the covalent bond $r_b$. }
	\label{fig07}
\end{figure*}
Finally, the pronounced asymmetry of the distribution $P(\sigma)$ with its elongate right wing is evidence of \textit{extended range of the correlation lengths} related to the short-range order of liquid gallium.  This inference explains  the distinct asymmetry of the main maximum of the radial distribution function $g(r)$ of liquid gallium (Fig.~\ref{fig03}). In Supplementary Information, we also show what information about the interaction of neighboring particles can be obtained from the distribution $P(\sigma)$.

\textbf{Extended range of the correlation lengths}. --
Let two particles be at a distance equal to a correlation length. The distance between a pair of particles is measured relative to their geometric centers. In the case of a crystal, this will be a distance between the corresponding sites of the crystal lattice; and this distance is directly defined by the lattice constants. In the case of a dense monatomic fluid, there are signs of short-range structural order with the corresponding well-defined correlation length. Herein, this correlation length characterizes an \textit{equilibrium arrangement} of the closest neighbors (their geometric centers) and is identified with the effective average particle size. This equilibrium mutual arrangement of a pair of neighboring particles is due to a compromise between the repulsion of particles at the short distances and the effective attraction of the particles of this pair caused by internal pressure in the system. Consequently, the correlation length associated with the short-range order can be evaluated even for the case of a dense fluid, where the particle interaction is purely repulsive. For the case of a dense fluid, this correlation length is a function of the internal pressure and/or the density.
Assuming that the particle interaction potential is spherically symmetric and the particle displacements near the distance associated with the equilibrium arrangement of the particles obey the Gaussian statistics, one obtains the simple Gaussian shape for the first peak of the radial distribution function, and this peak will be located at a distance equal to the correlation length~\cite{Frenkel_1975}. In polyvalent liquid metals like liquid gallium, the ion-ion interactions are not spherically symmetric and in these interactions there are no specified directions.  As a result, the short-range order will be characterized not by a single correlation length only, but by a range of the correlation lengths. To estimate this range for the concrete liquid, one can use the next simple method. Namely, one can expand the methodology commonly used to analyze the radial distribution function $g(r)$ of simple liquids and solve the optimization problem~\cite{Mokshin/Khusnutdinoff_JETP_2015}: How many Gaussian contributions must be used to reproduce the contour of the first peak of the radial distribution function in detail and with the necessary accuracy?  Then, the arrangement of these Gaussian functions will provide information on the inherent correlation lengths in the considered many-particle system.

For liquid gallium data, solving this optimization problem by means of the brute-force method [see Supplementary Information for details], we find that the contour of the first peak of $g(r)$ is correctly reproduced by a linear combination of the three Gaussian functions, whose positions determine the correlation range $[\sigma_1,\,\sigma_3]$ (see Fig.~\ref{fig07}). As it turned out, the positions of the Gaussians correspond exactly to the features of the contour of $g(r)$ that are also detected by the first derivative $dg(r)/dr$ [Figs.~\ref{fig07}(a) and (b)]. Remarkable findings follow from these \textit{approximate} results.  There is the range of the correlation lengths associated with the short-range order in liquid gallium. The width of this correlation range can be evaluated as $\sigma_3 - \sigma_1$, and it is about half angstrom. As seen in Fig.~\ref{fig07}(c), the quantities $\sigma_1$, $\sigma_2$, $\sigma_3$ as well as the width of the correlation range, $\sigma_3 - \sigma_1$, are practically independent on the temperature. The lower boundary $\sigma_1$ of the correlation range is comparable in values with the length $\sigma_0$ defined above, but it is larger than the covalent bond $r_b$. And, finally, the middle of the range $(\sigma_1+\sigma_3)/2 \simeq 2.83$~\AA~~approximates the known value of the average linear size of gallium atom~\cite{March_1990}.

Based on all these results, one can propose the following simplified \textit{semiquantitative} model for the structure that is realized in liquid gallium. The structural configuration formed by atoms (ions) of gallium is identified with the configuration generated by a close-packed system of \textit{soft quasi-spheres}, the surfaces of which are deformed and contain convexities and concavities [see Fig.~\ref{fig07}(a)]. There are two \textit{equilibrium} radii $\sigma_1/2$ and $\sigma_3/2$, which quantify the size and shape of these quasi-spheres and which correspond to the radius of the core and the radius of an enveloping sphere. Packaging of such the quasi-spheres yields the radial distribution function with more blurred  peaks for the second, third and fourth pseudo-coordination shells in comparison with the case of radial distribution function for simple liquids.
In particular, as follows from estimates given in Fig.~\ref{fig03}(b), the ratio between the width $r_{0/1}$ of the first pseudo-coordination shell and the width $r_{1/2}$ of the second shell is $r_{0/1}/r_{1/2}=0.83$ for the case of liquid gallium, whereas one has $r_{0/1}/r_{1/2}=0.66$ for simple liquids~\cite{JETP_Lett_2017_en}. There is also difference in correspondence between the peak locations in the radial distribution function $g(r)$ for liquid gallium and simple liquids, that is directly caused by the difference in the short-range order.  For example,  for simple liquids~\cite{JETP_Lett_2017_en}, location of the second peak is determined as $\simeq 1.88\; r_0$; recall that $r_0$ is the location of the first peak in $g(r)$ [see Fig.~\ref{fig03}(b)]. In liquid gallium, the extended short-range order leads to appearance of specific structural features relevant, in particular, to the second pseudo-coordination shell. According to the simplified model, the location of the second peak must be estimated as $\simeq 1.88(\sigma_1+\sigma_3)/2$, that gives the correct value, whereas relation $\simeq 1.88\; r_0$ underestimates this value.
Due to the structural features, the resonant $k$-components appear in the static structure factor $S(k)$ in a manner different from the case of simple liquids. Finally, the broadened maxima of the radial distribution function result in a shift of the spike of the main maximum of the static structure factor $S(k)$ to the range of small $k$.

The results of our study indicate directly that the so-called de Gennes narrowing phenomenon occurs in liquid gallium mainly due to the local structure features, rather than due to the collective dynamics of atoms (ions). This is fully consistent with the main findings of the recent paper~\cite{Wu_2018}, as well as with conclusions given~\cite{Bosio_1976} and \cite{Scopigno_Leonardo_2005}. In fact, the structure of liquid gallium including the short-range order with the spatial scales typical for the de Gennes narrowing can be reproduced by the soft quasi-spheres model. This is also in agreement with the recent findings in Ref.~\cite{Montfrooij_2016} that the ``hard-spheres''-like dynamics is not able to reproduce the density-dependent features of the de Gennes narrowing for the cases of liquid gallium and mercury.

A relationship between the peculiarities of the particle interaction potential and such the kinetic parameter as the fragility index $m$ was recently discussed in Ref.~\cite{Krausser_Zaccone_2015}, where it is found, in particular, that the metallic glass-formers with a steeper repulsive part of the interatomic interaction must be more fragile~\cite{MG_JETPLett_2019}. Anisotropy of interparticle interaction can characterize the expressed effective softness of this interaction. In this regard, one can expect that the more pronounced effects of softness of interparticle interaction are correlated with lower values of the fragility index $m$ for the polyvalent melts. On the other hand, according to our model of the close-packed soft quasi-spheres, the width of the first peak of the radial distribution function $g(r)$ is defined by the softness of the interatomic repulsion and by the contribution of thermal effects. This softness is quantified by the parameter $\xi$ (see the third part of Supplementary Information). Then, the dimensionless parameter $\xi/\sigma_0$ can be directly correlated with the steepness of the interatomic repulsion -- the parameter $\lambda$ introduced in Ref.~\cite{Krausser_Zaccone_2015}. Recall that $\sigma_0$ is the lower correlation length associated with the short-range order, and $\xi/\sigma_0 \in [0;\, 1)$.

\section{Concluding remarks and proposals}

Analysis of the local structure and evaluation of the time scales for the neighborhood of the particle pairs performed on the basis of AIMD simulation results for liquid gallium reveals the absence of  long-lived Ga$_2$ dimers and any stable local crystal-like domains. The significant effect of valence electrons on the ion-ion interaction is reflected in the fact that ion-ion interaction cannot be reproduced in the framework of any spherically symmetric model. The ion-ion interaction in gallium gives rise the extended short-range structural order in the dense liquid phase. In other words, there is no single length scale that can be identified directly from the form of the interaction potential with the particle size. The extended short-range order determines the structural features, which are manifested in the specific contours of the radial distribution function and the static structure factor.

\vskip 0.5 cm

To conclude, we suggest the next proposals, that can extend the given study:

\vskip 0.3 cm

(i) The given results can be extended to such polyvalent metal melts as liquid germanium, bismuth, tin, antimony, arsenic. It should be noted that, as known from X-ray diffraction data, these melts have the similar structural features as liquid gallium.

(ii) The simplified model of the close-packed system of soft quasi-spheres proposed in this study can be used to develop a new family of the effective ion-ion interaction potentials for liquid polyvalent metals. In particular, for the case of the EAM-type potentials~\cite{Inogamov_2019}, the information extractable by means of this model can be applied to construct the contour of its pair-wise contribution, its rigidity and softness, and to define its parameters that take into account anisotropy in the ion-ion interaction (see also  the third part of Supplementary Information in the paper).

(iii) In view of the recent results of studying three-particle correlations in liquids~\cite{Mokshin/Galimzyanov_Physa_2017,Kob_2019,GLM_JCG_2019,Khusnutdinoff_JCG_2019}, it would be useful to study the effect of three-particle correlations on the extended short-range order in liquid polyvalent metals.

(iv) The following questions associated with the outcomes of the recent paper \cite{Dyre_PRX_2016} can be quite naturally formulated: Is it possible to determine from the structural properties of a liquid that it is simple?
Is it possible to propose criteria defined only in terms of features of the radial distribution function and/or the static structure factor for determining whether the liquid is simple?

\begin{acknowledgments}
\noindent This work is supported by the Russian Science Foundation (project No. 19-12-00022). First-principle molecular dynamics simulations were performed by using the computational cluster of Kazan Federal University and the computational facilities of Joint Supercomputer Center of RAS.
\end{acknowledgments}

\section*{Supplementary Information}

\subsection*{Computational details}

We performed \textit{ab initio} molecular dynamics simulations of equilibrium liquid gallium  at various temperatures from the temperature range $T=[325; 3000]$~K along the isobar $p=1.0$ atm. These simulations were realized within the Car-Parrinello method  as implemented in the Quantum Espresso package~\cite{r1,r2,r3} with the exchange-correlation
functional taken within the local density approximation by the Perdew-Zunger parametrization~\cite{r4,r5}. The simulation cell had the shape of a parallelepiped and included  five hundred atoms of gallium. To avoid the size effects, we applied in our simulations the periodic boundary conditions, according to which the simulation cell is duplicated in all the directions.

The electronic states expanded in a plane-wave basis have been truncated at $35$~Ry. The time-step used to integrate numerically the equations of motion is $8.2682$~a.u. ($0.2$~fs). A fictitious mass of $500$~a.u. was assigned to the electronic degrees of freedom. To calculate the electronic structure it was used the Kohn-Sham formulation of DFT.  All the Car-Parrinello molecular dynamics simulations were done within the $NpT$-ensemble, with the pseudo-Hamiltonian being conserved within $10^{-5}$~au/ps. The temperature was controlled by means of the Nose-Hoover chain of thermostats~\cite{r7} applied to the ionic and electronic degrees of freedom.

\vskip 0.5cm

\subsection*{Fit of the radial distribution function by a linear combination of the Gaussian functions: The Brute-force method}

We apply the brute-force method to solve the optimization problem associated with estimating the parameters of the following expansion for the first peak of the radial distribution function over the Gaussian functions:
\begin{equation} \label{eq_exp}
g(r) = \sum_i \frac{n_i}{2\pi \xi_i^2} \exp \left ( - \frac{(r-\sigma_i)^2}{2\xi_i^2}  \right ).
\end{equation}
Here, $\sigma_i$ is the $i$th correlation length, $n_i$ is the $i$th coordination number, and $\xi_i$ is the effective dispersion, which determines the width of the $i$th Gaussian function. Applying relation~(\ref{eq_exp}) allows one to realize the idea that the dynamics of two particles separated by a distance comparable with a correlation length obeys the Gaussian statistics. Extending this idea we obtain that each peak of $g(r)$ is reproduced either by a single Gaussian or by a linear combination of the Gaussian functions. We perform a fit of the first peak of the `experimental' function $g(r)$ by relation~(\ref{eq_exp}) according to the routine based on the next rules. First, it is necessary to determine the minimum number of the Gaussian functions that allows one to reproduce the contour of $g(r)$ with the required accuracy $\varepsilon=0.003$. Second, trial values of the parameter $\sigma_i$ are taking from the range  restricted by the distance corresponding to the first nonzero value of $g(r)$ and by the distance $r_{1/2}$ (see Fig.~3). Trial values of both the parameters $\xi_i$ and $n_i$ are interrelated and these values are choosing so that the width and height of a model function $g(r)$ do not exceed values of the experimental function $g(r)$. Third, the fitting procedure starts with use of a single Gaussian, and then their number increases successively until the required accuracy is achieved.

\vskip 0.5cm

\subsection*{What information about interaction of neighboring particles can be obtained from distribution $P(\sigma)$ ?}

Appearance of the non-zero values of the distribution $P(\sigma)$ at the lengths $\sigma < \sigma_0$ display the effect of overcoming the repulsion of neighboring particles due to their thermal motion. The width $\xi$ of the left wing of the distribution $P(\sigma)$, as shown in Fig.~6, can be used to recover the features of particle repulsion at the short distances $r \leq \sigma_0$. Namely, assuming that the effective interaction $u(r)$ of particles in the neighborhood of the distance value $r=\sigma_0$ is spherically symmetric, we have
\begin{equation}
u(\sigma_0) - u(\sigma_0 - \xi) \simeq k_B T.
\end{equation}
Then, within the harmonic approximation
\begin{equation}
u(\sigma_0 - \xi) \simeq u(\sigma_0) - \frac{1}{2} \left . \frac{\partial^2 u(r)}{\partial r^2} \right |_{r=\sigma_0} \cdot \xi^2 + \mathcal{O}(\xi^4),
\end{equation}
we obtain
\begin{equation}
\xi^2 \simeq 2 k_B T \left [ \left . \frac{\partial^2 u(r)}{\partial r^2} \right |_{r=\sigma_0} \right ]^{-1},
\end{equation}
where $k_B$ is the Boltzmann constant.

\bibliographystyle{unsrt}

\end{document}